\documentclass[letterpaper,10 pt,conference]{ieeeconf}
\IEEEoverridecommandlockouts
\usepackage{cite}
\usepackage{amsmath,amssymb,amsfonts}
\usepackage{algorithmic}
\usepackage{graphicx}
\usepackage{textcomp}
\usepackage{xcolor}
\usepackage{placeins}
\usepackage[caption=false,font=footnotesize]{subfig}
\usepackage{mathrsfs}

\def\BibTeX{{\rm B\kern-.05em{\sc i\kern-.025em b}\kern-.08em
    T\kern-.1667em\lower.7ex\hbox{E}\kern-.125emX}}

\def\mline{\vrule width4pt height2.5pt depth -2pt}

\begin{document}

\title{Structured input--output analysis of oblique turbulent bands in Waleffe flow}

\author{Jino George$^{1}$ and Chang Liu$^{1}$%
\thanks{$^{1}$School of Mechanical, Aerospace, and Manufacturing Engineering, University of Connecticut, Storrs, CT 06269, USA
        {\tt\small \{jino.george, chang\_liu\}@uconn.edu}}%
}
\maketitle

\begin{abstract}
This work employs structured input--output analysis (SIOA) to study Waleffe flow. The SIOA framework employs structured uncertainty to include the componentwise structure of nonlinearity in Navier-Stokes equations, and SIOA quantifies the flow response using structured singular values. The structured input--output analysis identifies the wavelength and inclination angle of oblique turbulent bands observed in large-domain direct numerical simulations. The structured input--output response scales over Reynolds number as $\sim Re^{1.7}$.

\end{abstract}

\vspace{-1ex}
\section{Introduction}
\vspace{-1ex}

Structured singular value \cite{packard1993complex} from robust control theory has recently been introduced to analyze transition to turbulence of wall-bounded shear flows \cite{liu2021structured,liu2022structured,shuai2023structured}. This framework, structured input--output analysis (SIOA), employs structured uncertainty to capture a key nonlinear mechanism that weakens the linear lift-up mechanism \cite{liu2021structured}. The inclusion of nonlinear properties in SIOA has allowed us to rapidly predict transitional flow structures that typically require computationally expensive direct numerical simulations (DNS). For example, SIOA identified oblique turbulent--laminar patterns observed in DNS and experiments with a very large domain \cite{prigent2003long,duguet2010formation,klotz2021experimental,taylor2016new}, including plane Couette flow (PCF), plane Poiseuille flow \cite{liu2021structured}, stratified PCF \cite{liu2022structured}, and Couette-Poiseuille flow \cite{shuai2023structured}. 

This work will employ SIOA in a new flow configuration: Waleffe flow \cite{waleffe1997self}, which was originally proposed to analyze self-sustaining processes in wall-bounded shear flows without the boundary layer. The Waleffe flow was shown to have several similarities to other canonical flows \cite{chantry2016turbulent}, e.g., oblique turbulent bands have also been observed in Waleffe flow \cite{chantry2016turbulent}. The goal of this work is to investigate whether SIOA can predict oblique turbulent bands in Waleffe flow, and also use SIOA to compare Waleffe flow with other canonical flows. 

\vspace{-1ex}
\section{Structured input--output analysis}
\vspace{-1ex}

\begin{figure}[t!]
    \centering
    \includegraphics[width=0.5\linewidth]{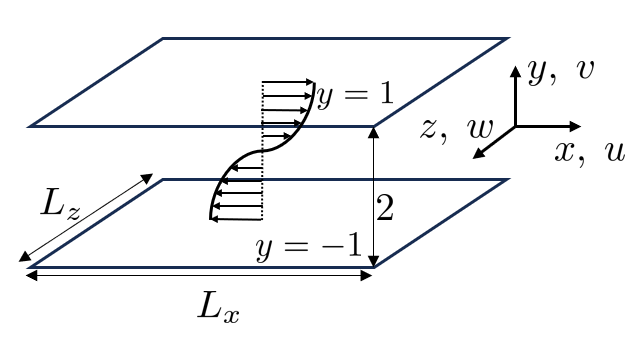}
    \vspace{-3ex}
    \caption{Flow configuration of a sinusoidal shear flow.}
    \label{fig:domain waleffe}
    \vspace{-5ex}
\end{figure}

We consider the Waleffe flow  \cite{waleffe1997self,chantry2016turbulent} shown in Fig. \ref{fig:domain waleffe}, describing flow between two parallel plates driven by a sinusoidal body force with stress-free boundary conditions at the wall. This flow describes the self-sustaining process in wall-bounded shear flows. The velocity vector is denoted as $\boldsymbol{u} = [u,v,w]^\text{T}$, where $u$, $v$, $w$ are velocity components in streamwise, wall-normal, and spanwise directions. The length variables are normalized by the channel half-height $h$, leading to the top and bottom walls as $y=\pm 1$. The velocity is normalized by the wall velocity $U_0$, and the time is normalized by $h/U_0$. The Reynolds number is defined as $Re = \frac{U_0  h}{\nu}$, where $\nu$ is the kinematic viscosity. Here, we consider laminar base flow $\boldsymbol{U}(y)=U(y)\boldsymbol{e}_x=\sin(\pi y/2)\boldsymbol{e}_x$ driven by the body force $\boldsymbol{F}=-\nabla^2\boldsymbol{U}/Re$ \cite{waleffe1997self,chantry2016turbulent}. We then decompose the velocity as  $\boldsymbol{u}_{\text{tot}}=\boldsymbol{U}(y)+\boldsymbol{u}$ and decompose pressure as $p_{\text{tot}}=P+p$, leading to governing equations:
\begin{subequations}
\label{eq:ns_decomposed}
\begin{align}
\partial_t \boldsymbol{u}
+ \boldsymbol{U}{\cdot}\nabla \boldsymbol{u}+\boldsymbol{u}{\cdot}\nabla \boldsymbol{U}
=& -\nabla p
{+} \nabla^{2}\boldsymbol{u}/Re-\boldsymbol{u}{\cdot}\nabla\boldsymbol{u}, \;
\label{eq:ns_decomposed_momentum}\\
\nabla {\cdot}\boldsymbol{u}=&\;0
\label{eq:divergence}
\end{align}
\end{subequations}
with boundary conditions $\frac{du}{dy}(y=\pm 1)=\frac{dw}{dy}(y=\pm 1)=0$ and $v(y=\pm 1)=0$. We then model the nonlinear term $-\boldsymbol{u}{\cdot}{\nabla}\boldsymbol{u}$ in Eq. \eqref{eq:ns_decomposed_momentum} as a structured forcing model \cite{liu2021structured}:
\begin{equation}\label{SIOAForcing}
\begin{split}
    \textit{\textbf{f}}_\xi :=-\boldsymbol{u}_\xi{\cdot}{\nabla}\boldsymbol{u}
     = \begin{bmatrix}
-\boldsymbol{u}_\xi{\cdot}{\nabla}u\\
-\boldsymbol{u}_\xi{\cdot}{\nabla}v\\
-\boldsymbol{u}_\xi{\cdot}{\nabla}w
\end{bmatrix}=:\begin{bmatrix}
\textit{f}_{x,\xi}\\
\textit{f}_{y,\xi}\\
\textit{f}_{z,\xi}
\end{bmatrix}
\end{split}
\end{equation}
with the gain operator $-\boldsymbol{u}_\xi$ invariant in $x$, $z$, and $t$. This forcing model preserves the componentwise structure of the nonlinear term $-\boldsymbol{u}{\cdot}{\nabla} \boldsymbol{u}$ \cite{liu2021structured}, such that each component of the forcing is mostly influenced by that component of velocity. 

We then perform the Fourier transform $
\widehat{\psi}(y;k_x,k_z,\omega)
:=
\iiint_{-\infty}^{\infty}
\psi(x,y,z,t) 
e^{-\text{i}(k_x x + k_z z + \omega t)}
\, dx\,dz\,dt$ and rewrite Eq. \eqref{eq:ns_decomposed} as $\text{i}\omega\mathcal{E}\boldsymbol{\Psi} = \mathcal{A}\boldsymbol{\Psi} + \mathcal{B}\widehat{\boldsymbol{f}}_\xi,\;\;\widehat{\boldsymbol{u}} = \mathcal{C}\,\boldsymbol{\Psi},$ where $\boldsymbol{\Psi} = [\widehat{u}, \widehat{v}, \widehat{w}, \widehat{p}]^\mathrm{T}$ is the state variable vector, and the output is $\widehat{\boldsymbol{u}}=[\widehat{u}, \widehat{v}, \widehat{w}]^\text{T}$. The $\mathcal{A}$ operator is
\begin{equation}\mathcal{A} = 
\begin{bmatrix}
\mathcal{A}_{11} & -\partial_y U & 0 & -\text{i}k_x \\
0 & \mathcal{A}_{11} & 0 & -\partial_y \\
0 & 0 & \mathcal{A}_{11} & -\text{i}k_z \\
\text{i}k_x & \partial_y & \text{i}k_z & 0
\end{bmatrix},
\label{eq:Streamwise_A}
\end{equation} 
where $
\mathcal{A}_{11} = -\text{i}k_x U  +  \widehat{\nabla}^2/Re$ and $\widehat{\nabla}^2=\partial_{yy}-k_x^2-k_z^2$. The $\mathcal{B}$, $\mathcal{C}$, and $\mathcal{E}$ operators are defined as $\mathcal{B}=\begin{bmatrix}\mathcal{I}_{3\times 3} \\ 0_{1\times 3}\end{bmatrix}$, $\mathcal{C}=\begin{bmatrix}\mathcal{I}_{3\times 3}  & 0_{3\times 1}\end{bmatrix}$, and  $\mathcal{E}=\text{diag}(\mathcal{I}_{3\times 3},0)$. Here, $\text{diag}(\cdot)$ means block diagonal, $\mathcal{I}_{3\times 3}$ is an identity matrix, and ${0}_{1\times 3}$ and $0_{3\times 1}$ are zero matrices with size indicated in the subscript.

We can then obtain the spatio-temporal frequency response operator $\mathcal{H}$ that maps the input forcing term $\boldsymbol{\widehat{f}}_\xi$ to the output velocity vector at each spatiotemporal frequency: $\widehat{\boldsymbol{u}}(y;k_x,k_z,\omega) = \mathcal{H}\,\widehat{\boldsymbol{f}}_\xi(y;k_x,k_z, \omega)$ with $\mathcal{H}(y;k_x,k_z,\omega) = \mathcal{C} (\text{i} \omega \mathcal{E} - \mathcal{A})^{-1} \mathcal{B}$. To isolate the $\boldsymbol{u}_\xi$, we decompose the structured forcing as \cite{liu2021structured}:
\begin{equation}\label{RSIOAForcing}
    \begin{bmatrix}
        \widehat{\textit{f}}_{x,\xi} \\ \widehat{\textit{f}}_{y,\xi} \\ \widehat{\textit{f}}_{z,\xi}
    \end{bmatrix} =   \underbrace{\begin{bmatrix}\widehat{\boldsymbol{u}}^T_{\xi} & & 
 \\
& \widehat{\boldsymbol{u}}^T_{\xi} &\\
 & & \widehat{\boldsymbol{u}}^T_{\xi}
\end{bmatrix}}_{\widehat{\boldsymbol{u}}_{\Xi}}
\;\text{diag}\left(\widehat{{\nabla}},\widehat{{\nabla}},\widehat{{\nabla}}\right)\begin{bmatrix}
        \widehat{{u}}\\ \widehat{{v}}\\ \widehat{{w}}
    \end{bmatrix},
    \end{equation}
 where the block diagonal  $\widehat{\boldsymbol{u}}_{\Xi}$ is to be computed by structured singular value. We combine the gradient operator with $\mathcal{H}$ to obtain 
$\mathcal{H}_{{\nabla}}:=\text{diag}\left(\widehat{{\nabla}},\widehat{{\nabla}},\widehat{{\nabla}}\right) \mathcal{H},$ leading to a feedback interconnection between $\mathcal{H}_{{\nabla}}$ and $\widehat{\boldsymbol{u}}_{\Xi}$ \cite{liu2021structured}.

We then denote the numerical discretization of $\mathcal{H}_\nabla$ as $\mathbf{H}_{\nabla}$ and the discretization of $\widehat{\boldsymbol{u}}_\xi^\text{T}$ as $\widehat{\mathbf{u}}_\xi^\text{T}$, allowing us to define a set of structured uncertainty as $
\widehat{\mathbf{U}}_{\Xi}:=\{\text{diag}(-\widehat{\mathbf{u}}_{\xi}^{\text{T}},-\widehat{\mathbf{u}}_{\xi}^{\text{T}},-\widehat{\mathbf{u}}_{\xi}^{\text{T}}):-\widehat{\mathbf{u}}_{\xi}^{\text{T}}\in \mathbb{C}^{N_y \times 3N_y}\}$. Here, $N_y$ denotes the number of Chebyshev collocation points in the $y$ direction. The structured singular value is then defined as \cite{packard1993complex}: $\mu_{\widehat{\mathbf{U}}_{\Xi}}\left[\mathbf{H}_{\nabla }\right]=1/{\text{min}\{\bar{\sigma}[\widehat{\mathbf{u}}_{\Xi}]:\widehat{\mathbf{u}}}_{\Xi}{\in}\widehat{\mathbf{U}}_{\Xi},\text{det}[\mathbf{I}-\mathbf{H}_{\nabla}\widehat{\mathbf{u}}_{\Xi}]=0\}$ with
$\mu_{\widehat{\mathbf{U}}_{\Xi}}\left[\mathbf{H}_{\nabla }\right]= 0$ 
if $\forall\,\,\widehat{\mathbf{u}}_{\Xi}\in\widehat{\mathbf{U}}_{\Xi}$, $\text{det}[\mathbf{I}-\mathbf{H}_{\nabla}\widehat{\mathbf{u}}_{\Xi}]\neq0$. Here, $\bar{\sigma}[\cdot]$ is the largest singular value. We compute $\mu$ using \texttt{mussv} package in MATLAB, relaxing $\widehat{\mathbf{U}}_{\Xi}$ as non-repeated blocks. We then define the structured input--output response as $
\|\mathcal{H}_{{\nabla} }\|_{\mu}(k_x,k_z):=
\sup\limits_{\omega \in \mathbb{R}}\mu_{\widehat{\mathbf{U}}_{\Xi}}
\left[\mathbf{H}_{{\nabla}}(y;k_x,k_z,\omega)\right]$, and the unstructured input--output response ($\mathcal{H}_\infty$ norm)  \cite{jovanovic2021bypass} as $
\|\mathcal{H}\|_{\infty}(k_x,k_z):=
\sup\limits_{\omega \in \mathbb{R}}\bar{\sigma}
\left[\mathbf{H}(y;k_x,k_z,\omega)\right]$. 

\FloatBarrier

\vspace{-1ex}
\section{Results and discussions}
\vspace{-1ex}

\begin{figure}[t]
    \centering

    \subfloat[]{%
        \includegraphics[width=0.48\linewidth]{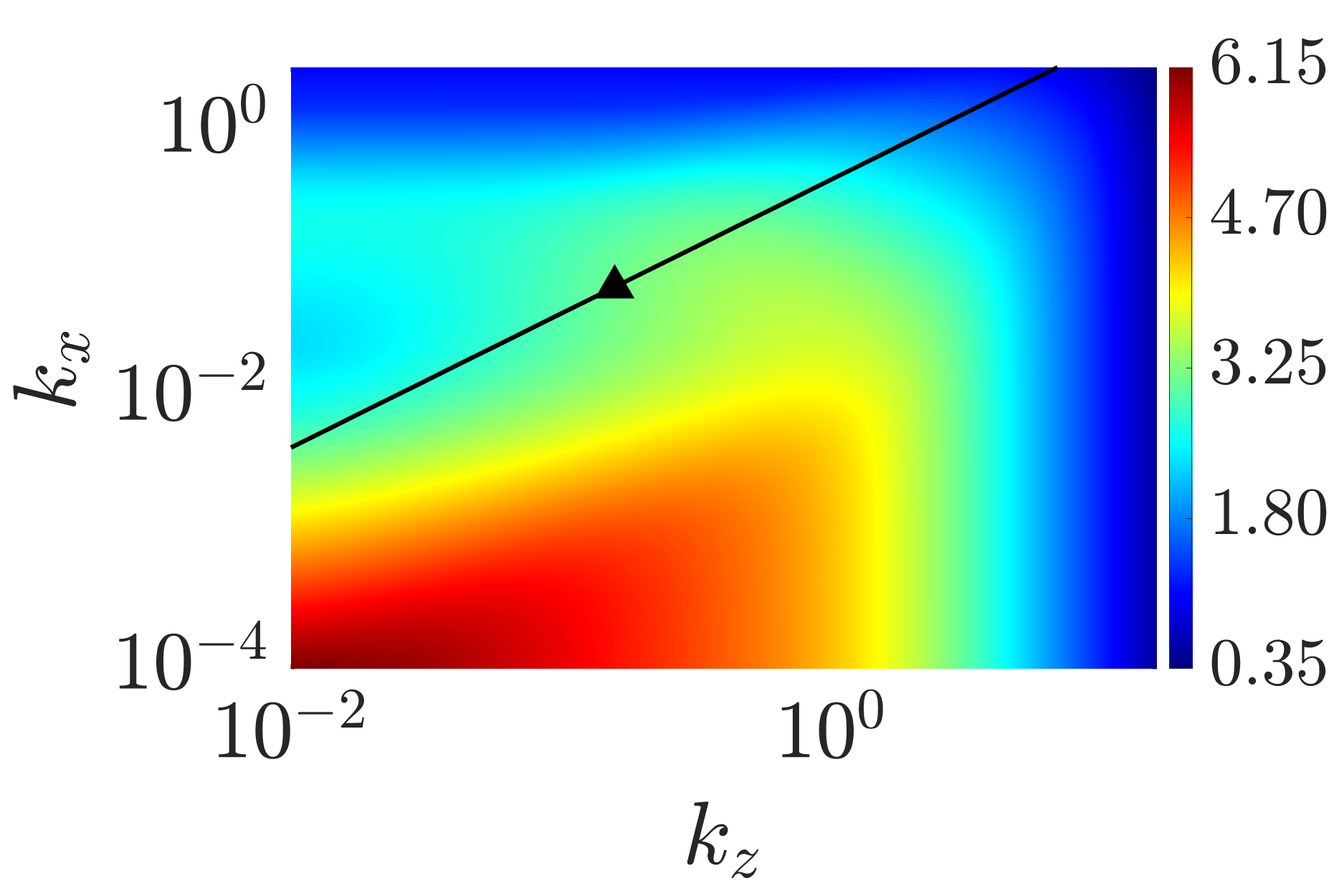}%
        \label{fig:hinf_re350}
    }
    \hfill
    \subfloat[]{%
        \includegraphics[width=0.48\linewidth]{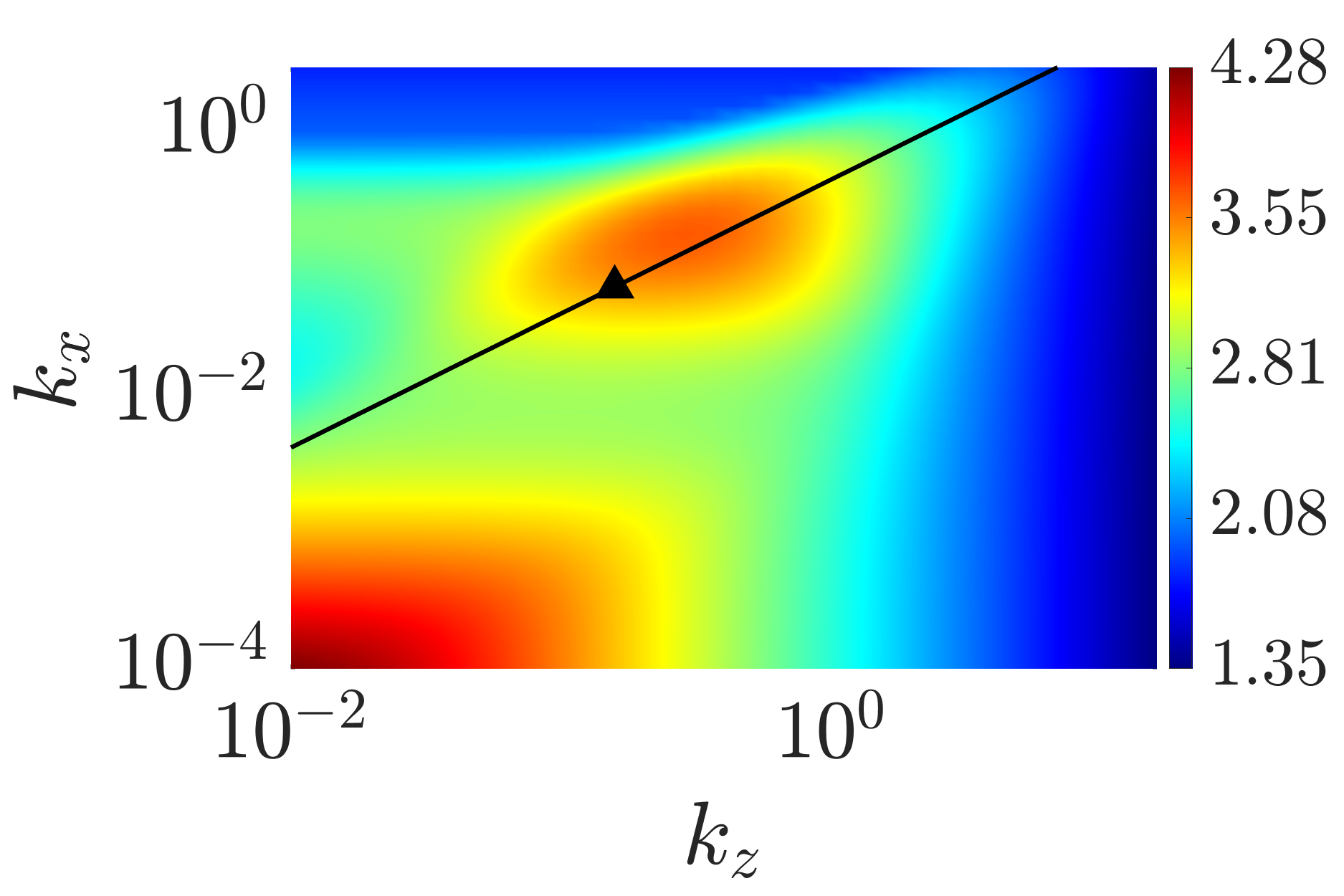}%
        \label{fig:hinf_re2000}
    }
\vspace{-2ex}
    \subfloat[]{%
        \includegraphics[width=0.48\linewidth]{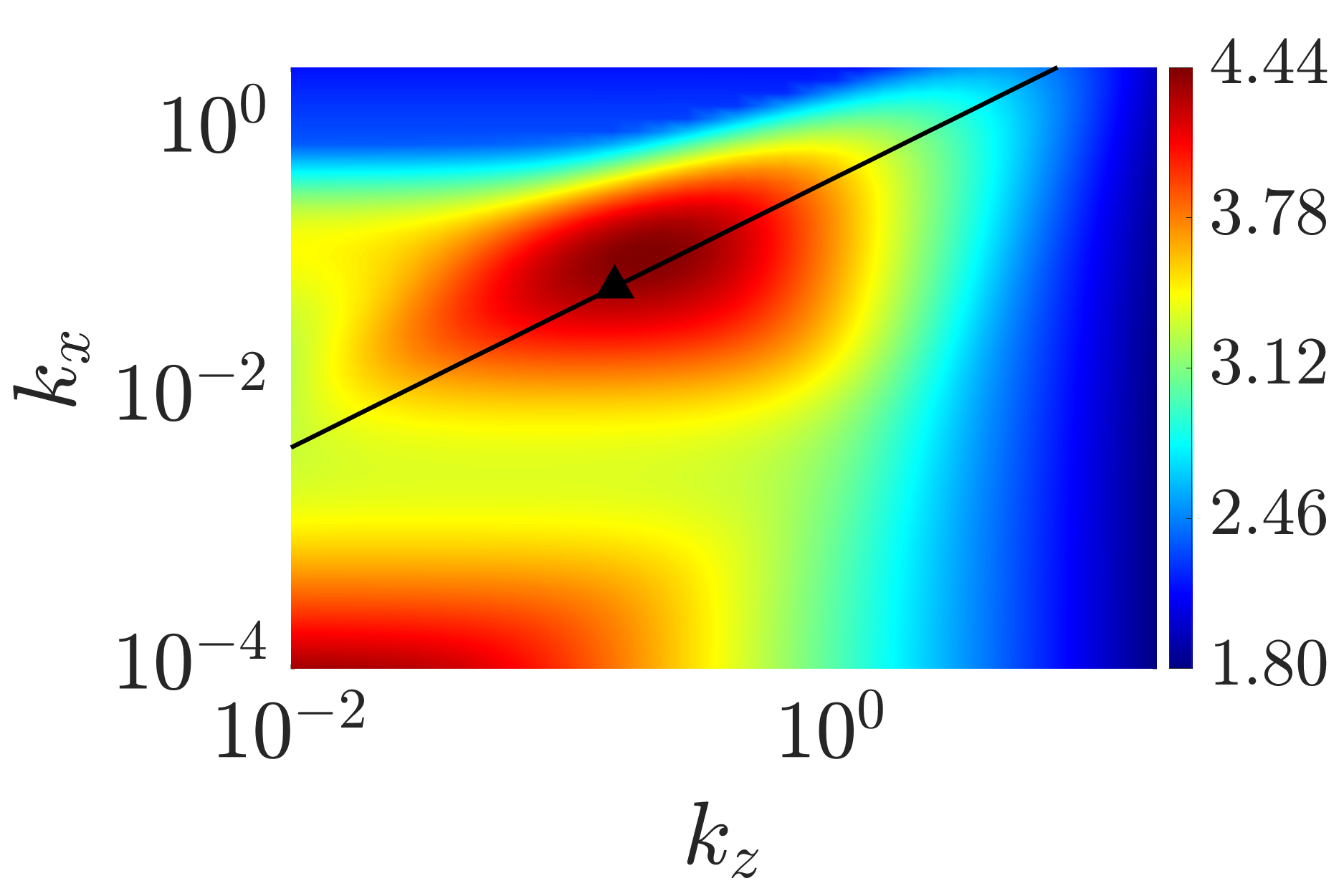}%
        \label{fig:mumax_re350}
    }
    \hfill
    \subfloat[]{%
        \includegraphics[width=0.48\linewidth]{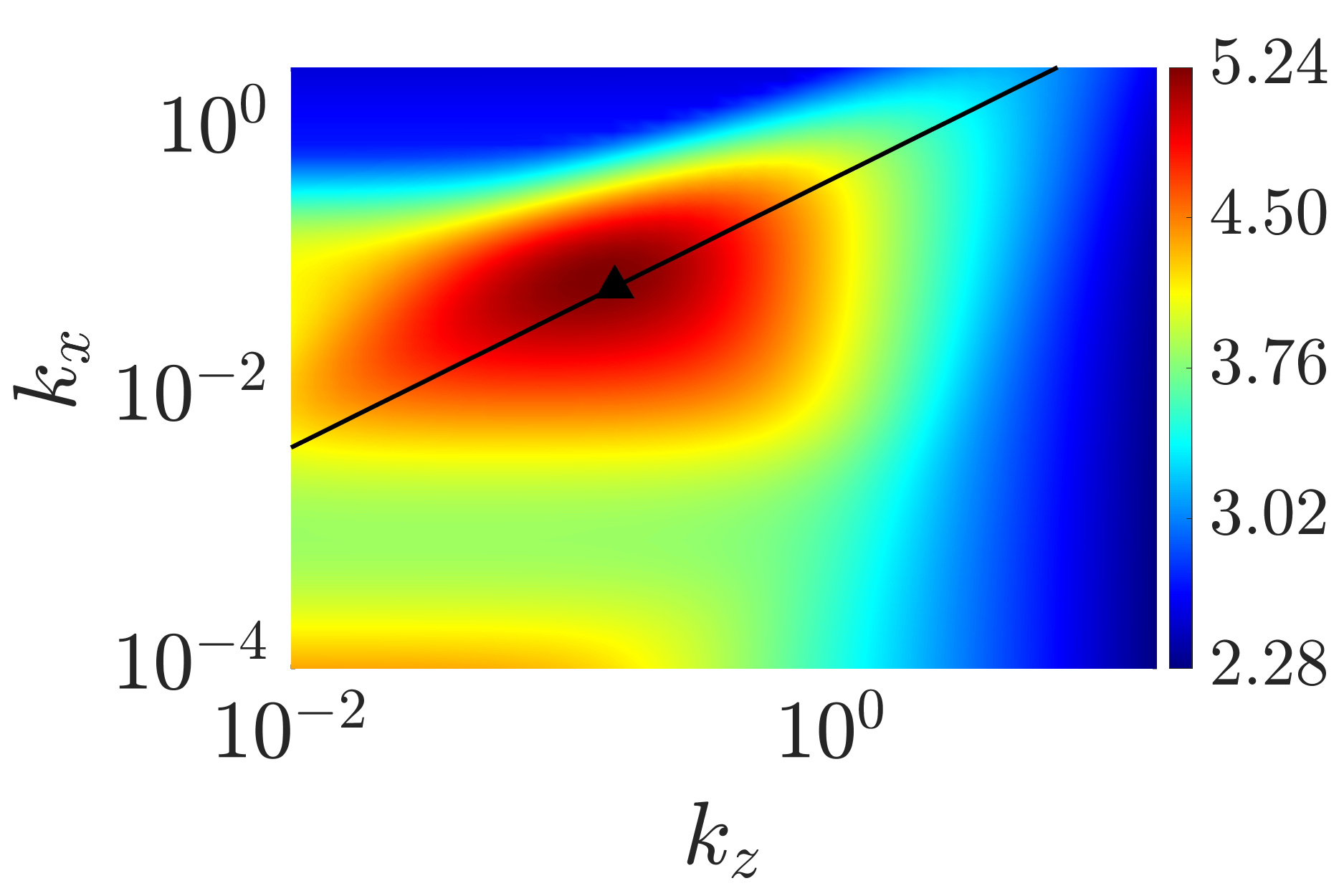}%
        \label{fig:mumax_re2000}
    }
    \caption{(a) $\log_{10}\!\left[\|\mathcal{H}\|_{\infty}(k_x,k_z)\right]$ at $Re=350$. Panels (b-d) are $\log_{10}\!\left[\|\mathcal{H}_{\nabla}\|_{\mu}(k_x,k_z)\right]$ at $Re = 350$, $ 1000$, and $3000$, respectively, with wavelength $(\lambda_x = 90,\lambda_z = 40, \blacktriangle)$ and inclination angle $\theta=23.96^\circ =\text{tan}^{-1}(\lambda_z/\lambda_x)$ ($\mline\mline$) of the oblique turbulent band at $Re = 350$ \cite{chantry2016turbulent}.}
    \label{fig:hinf_mumax_re350_re2000}
    \vspace{-2ex}
\end{figure}

Figure \ref{fig:hinf_mumax_re350_re2000}(a) shows (unstructured) input--output response based on $\|\mathcal{H}\|_\infty$ displaying a peak at $k_x\approx 0$ and $k_z\approx 0$. However, structured input--output response based on $\|\mathcal{H}_\nabla\|_\mu$ in Fig. \ref{fig:hinf_mumax_re350_re2000}(b) shows a local peak between $k_x\in[10^{-2},1]$ and $k_z\in [10^{-1},1]$ with the shape of the peak region resembles that of plane Couette flow \cite{liu2021structured}. The (unstructured) input--output analysis $\|\mathcal{H}\|_\infty$ in Fig. \ref{fig:hinf_mumax_re350_re2000}(a) does not show a high amplification near the wavelength ($\blacktriangle$) and inclination ($\mline\mline$) angle of oblique turbulent bands, while the structured input--output analysis in Fig. \ref{fig:hinf_mumax_re350_re2000}(b) successfully captures oblique turbulent bands in Waleffe flow \cite{chantry2016turbulent}. The SIOA captures oblique flow structures because it preserves the componentwise structure of nonlinear terms, weakening the streamwise elongated structures driven by the lift-up mechanism \cite{liu2021structured}.

\begin{figure}[t]
    \centering
    \subfloat[$k_x^{M}$ and $k_z^{M}$ versus \ $Re$]{%
        \includegraphics[width=0.48\linewidth]{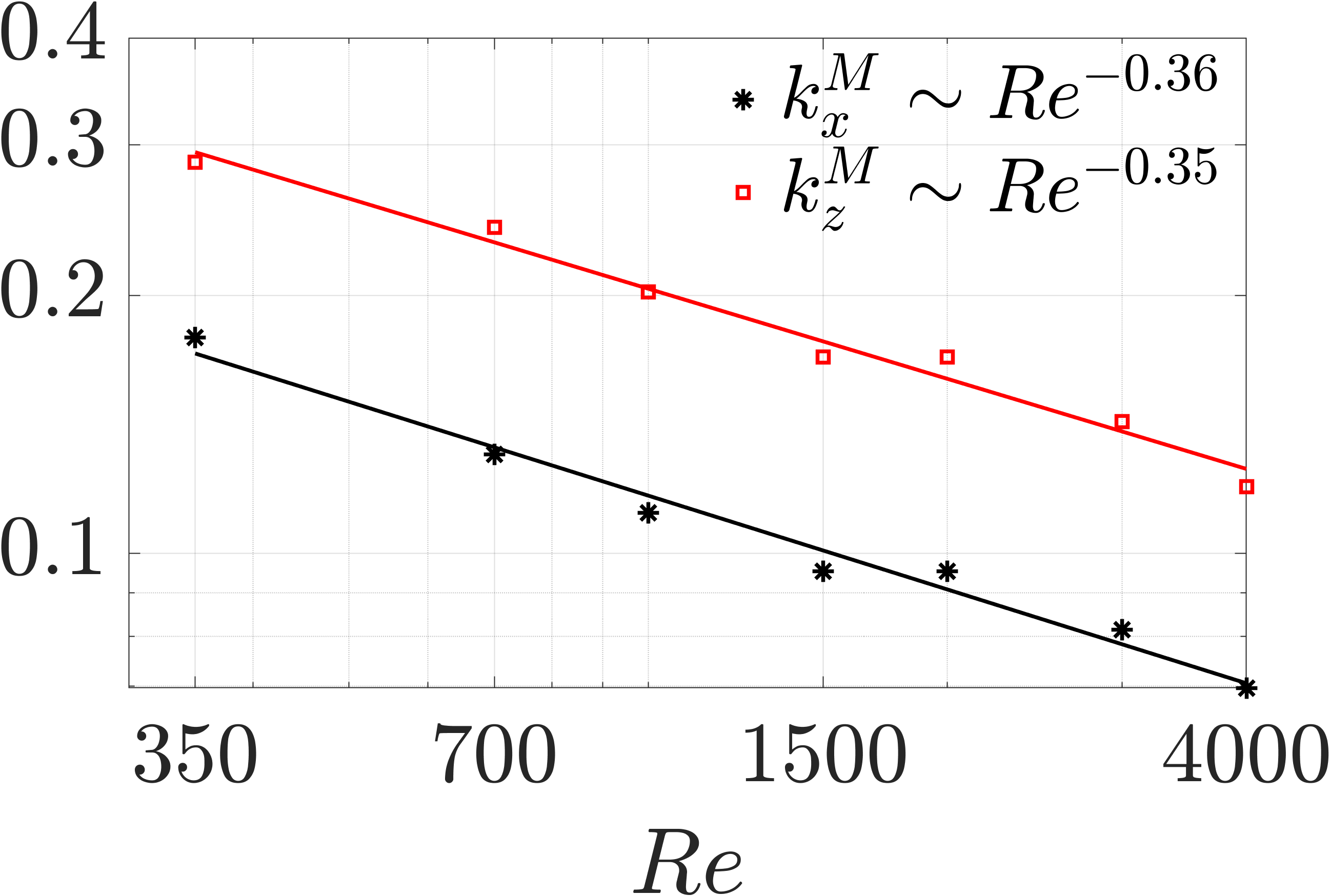}%
        \label{fig:kxkz_vs_Re}
    }
\hfill 
\subfloat[$\|\mathcal{H}_{\nabla}\|_{\mu}^{M}$ versus \ $Re$]{%
        \includegraphics[width=0.48\linewidth]{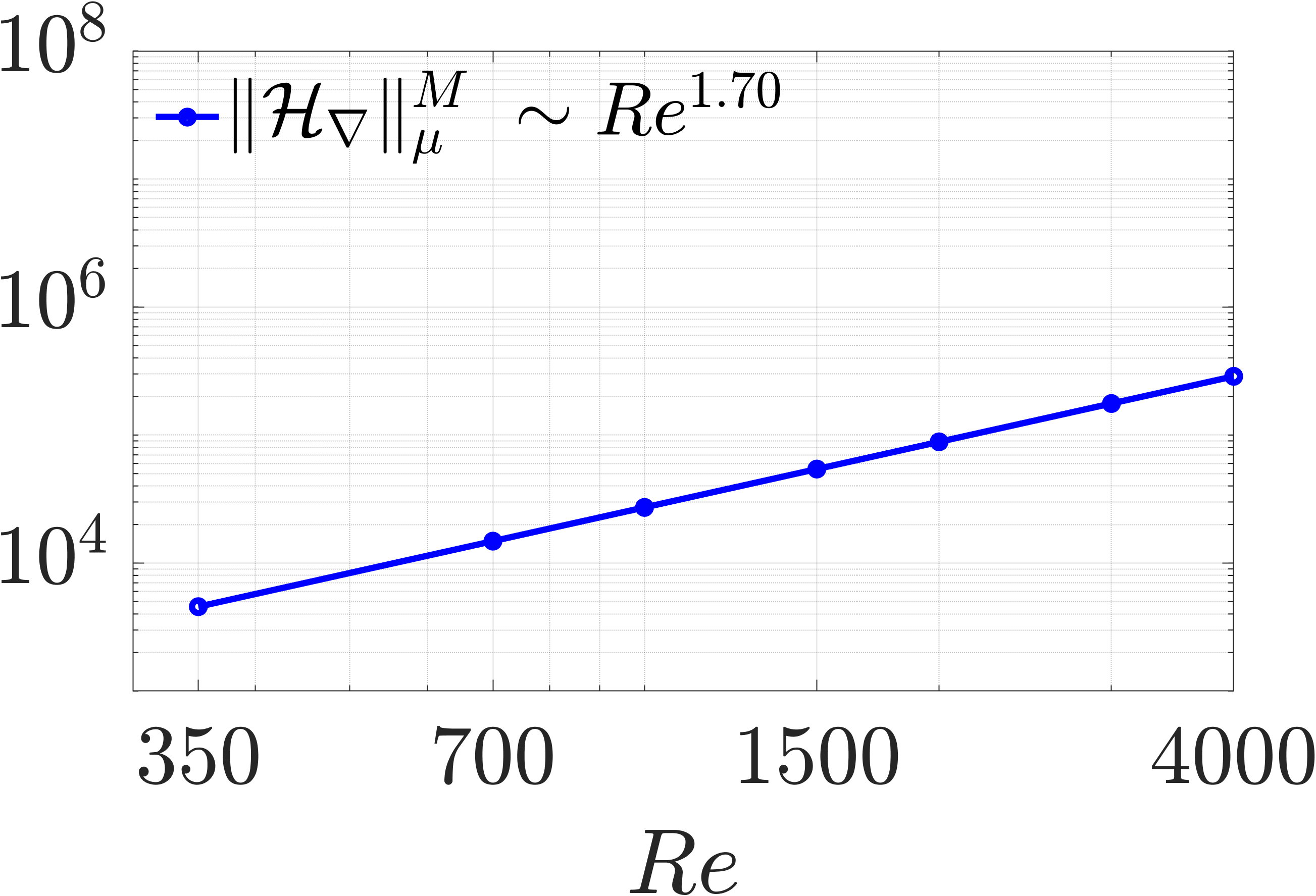}%
        \label{fig:Hmu_vs_Re}
    }
    \caption{Reynolds number dependence of (a) $(k_x^{M},k_z^{M}):=\mathop{\underset{k_x,k_z}{\arg\max}}\|\mathcal{H}_{\nabla}\|_{\mu}(k_x,k_z)$ and (b) $\|\mathcal{H}_{\nabla}\|_{\mu}^{M}:=\underset{k_x,k_z}{\max}\|\mathcal{H}_{\nabla}\|_{\mu}(k_x,k_z)$ within the wavenumber range $k_x \geq 10^{-3}$ and $k_z \geq 10^{-2}$.}
    \label{fig:Hmu_and_kxkz_vs_Re}
    \vspace{-3ex}
\end{figure}

As the Reynolds number increases, peak region of $\|\mathcal{H}_\nabla\|_\mu(k_x,k_z)$ in Figs. \ref{fig:hinf_mumax_re350_re2000}(b-c) shifts to smaller $k_x$ and $k_z$ (Fig. \ref{fig:Hmu_and_kxkz_vs_Re}a) with scaling $k_x^M\sim Re^{-0.36}$ and $k_z^M\sim Re^{-0.35}$ close to Couette-Poiseuille flow \cite{shuai2023structured}. The peak regions of $\|\mathcal{H}_\nabla\|_\mu(k_x,k_z)$ in Figs. \ref{fig:hinf_mumax_re350_re2000}(b-c) still pass through ($\blacktriangle$) and ($\mline\mline$) representing oblique turbulent bands at $Re=350$ \cite{chantry2016turbulent}. The Reynolds number scaling in Fig. \ref{fig:Hmu_and_kxkz_vs_Re}(b) shows that $\|\mathcal{H}_\nabla\|_\mu^M\sim Re^{1.7}$ over $Re\in [350,4000]$, which increases faster over Reynolds number compared with plane Couette flow $\|\mathcal{H}_\nabla\|_\mu^M\sim Re^{1.1}$, plane Poiseuille flow $\|\mathcal{H}_\nabla\|_\mu^M\sim Re^{1.5}$ \cite{liu2021structured}, and Couette-Poiseuille flow \cite{shuai2023structured}. This difference in Reynolds number scaling suggests that Waleffe flow may show different flow features at high Reynolds numbers.

\vspace{-1ex}
\section{Conclusion}
\vspace{-1ex}

This work shows that SIOA can effectively capture oblique turbulent bands observed in DNS of Waleffe flow \cite{chantry2016turbulent}, similar to its previous success in other shear flows \cite{liu2021structured,liu2022structured,shuai2023structured}. As Reynolds number increases, the horizontal wavenumber leading to the maximal structured input--output response decreases, and the structured input--output response increases as $\|\mathcal{H}_{\nabla}\|_{\mu}^{M} \sim $  $Re^{1.7}$ with a scaling exponent larger than that of plane Couette flow and plane Poiseuille flow \cite{liu2021structured}. 

\vspace{-1ex}

\bibliographystyle{IEEEtran}
\bibliography{references}

@article{klotz2021experimental,
  title={Experimental measurements in plane {C}ouette--{P}oiseuille flow: dynamics of the large-and small-scale flow},
  author={Klotz, Lukasz and Pavlenko, AM and Wesfreid, JE},
  journal={J. Fluid Mech.},
  volume={912},
  pages={A24},
  year={2021},
  publisher={Cambridge University Press}
}

@article{taylor2016new,
  title={A new method for isolating turbulent states in transitional stratified plane {C}ouette flow},
  author={Taylor, J R and Deusebio, Enrico and Caulfield, C P and Kerswell, Richard R},
  journal={J. Fluid Mech.},
  volume={808},
  pages={R1},
  year={2016},
  publisher={Cambridge University Press}
}

@article{liu2021structured,
  title={Structured input--output analysis of transitional wall-bounded flows},
  author={Liu, Chang and Gayme, Dennice F},
  journal={J. Fluid Mech.},
  volume={927},
  pages={A25},
  year={2021},
  publisher={Cambridge University Press}
}

@article{shuai2023structured,
  title={Structured input--output analysis of oblique laminar--turbulent patterns in plane {C}ouette--{P}oiseuille flow},
  author={Shuai, Yu and Liu, Chang and Gayme, Dennice F},
  journal={Int. J. Heat Fluid Flow},
  volume={103},
  pages={109207},
  year={2023},
  publisher={Elsevier}
}

@article{chantry2016turbulent,
  title={Turbulent--laminar patterns in shear flows without walls},
  author={Chantry, Matthew and Tuckerman, Laurette S and Barkley, Dwight},
  journal={J. Fluid Mech.},
  volume={791},
  pages={R8},
  year={2016},
  publisher={Cambridge University Press}
}

@article{waleffe1997self,
  title={On a self-sustaining process in shear flows},
  author={Waleffe, Fabian},
  journal={Phys. Fluids},
  volume={9},
  pages={883--900},
  year={1997},
  publisher={American Institute of Physics}
}

@article{prigent2003long,
  title={Long-wavelength modulation of turbulent shear flows},
  author={Prigent, Arnaud and Gr{\'e}goire, Guillaume and Chat{\'e}, Hugues and Dauchot, Olivier},
  journal={Physica D},
  volume={174},
  pages={100--113},
  year={2003},
  publisher={Elsevier}
}

@article{duguet2010formation,
  title={Formation of turbulent patterns near the onset of transition in plane {C}ouette flow},
  author={Duguet, Yohann and Schlatter, Philipp and Henningson, Dan S},
  journal={J. Fluid Mech.},
  volume={650},
  pages={119--129},
  year={2010},
  publisher={Cambridge University Press}
}

@article{liu2022structured,
  title={Structured input--output analysis of stably stratified plane {C}ouette flow},
  author={Liu, Chang and Caulfield, C P and Gayme, Dennice F},
  journal={J. Fluid Mech.},
  volume={948},
  pages={A10},
  year={2022},
  publisher={Cambridge University Press}
}

@article{jovanovic2021bypass,
  title={From bypass transition to flow control and data-driven turbulence modeling: an input--output viewpoint},
  author={Jovanovi{\'c}, Mihailo R},
  journal={Annu. Rev. Fluid Mech.},
  volume={53},
  pages={311--345},
  year={2021},
  publisher={Annual Reviews}
}

@article{packard1993complex,
  title={The complex structured singular value},
  author={Packard, Andrew and Doyle, John},
  journal={Automatica},
  volume={29},
  pages={71--109},
  year={1993},
  publisher={Elsevier}
}
\vspace{12pt}

\end{document}